\documentclass[a4paper,12pt]{article}
\usepackage{graphicx}
\usepackage[footnotesize]{caption}

\usepackage{multirow}
\usepackage{dsfont}
\usepackage{amssymb}
\usepackage{slashed}
\usepackage{mathtools}
\usepackage{nicefrac}

\usepackage{axodraw}

\newcommand{\rr}[1]{\mathrm{#1}}
\newcommand{\cc}[1]{\mathcal{#1}}

\newcommand{\dd}[1]{\mathds{#1}}

\newcommand{\be}{\begin{eqnarray}}
\newcommand{\nn}{\nonumber}
\newcommand{\ee}{\end{eqnarray}}
\newcommand{\ba}{\begin{array}}
\newcommand{\ea}{\end{array}}

\newcommand{\essm}{$E_6$SSM }
\newcommand{\essmSTOP}{$E_6$SSM. }
\newcommand{\ezssm}{EZSSM }
\newcommand{\ezssmSTOP}{EZSSM. }
\newcommand{\neff}{N_{\rr{eff}}}
\newcommand{\eff}{{\rr{eff}}}
\newcommand{\eq}{{\rr{eq}}}

\begin{document}

\begin{titlepage}

\begin{center}
{\sffamily\Large Bino Dark Matter and Big Bang Nucleosynthesis 
in the Constrained $E_6$SSM 
with Massless Inert Singlinos
}
\\[8mm]
Jonathan~P.~Hall and Stephen~F.~King
\\[3mm]
{\small\it
School of Physics and Astronomy, University of Southampton,\\
Southampton, SO17 1BJ, UK}
\\[1mm]
\end{center}
\vspace*{20mm}

\begin{abstract}
\noindent
We discuss a new variant of the $E_6$ inspired supersymmetric standard
model ($E_6$SSM) in which the two inert singlinos are exactly massless and 
the dark matter candidate has a dominant bino component.
A successful relic density is achieved via
a novel mechanism in which the bino scatters inelastically 
into heavier inert Higgsinos during the time of thermal freeze-out.
The two massless inert singlinos contribute to the
effective number of neutrino species at the time of Big Bang Nucleosynthesis, 
where the precise contribution depends
on the mass of the $Z'$ which keeps them in equilibrium. For example for 
$m_{Z'} > 1300$~GeV we find $N_\eff \approx  3.2$, where the smallness of the
additional contribution is due to entropy dilution. 
We study a few benchmark points
in the constrained $E_6$SSM with massless inert singlinos to illustrate this
new scenario.
\end{abstract}

\end{titlepage}
\newpage
\setcounter{footnote}{0}

\section{Introduction}
The evidence for dark matter is now very strong. Initially proposed to allow
an explanation of observed galactic rotation curves, we now also see
its effects in the cosmic microwave background (CMB).
The CMB baryon acoustic oscillation measurements from WMAP allow us to estimate
the relative density of cold dark matter in the universe to be
$\Omega_{\rr{CDM}} = (0.1099 \pm 0.0062) / h^2$~\cite{Dunkley2009},
where $h$ is the reduced Hubble parameter $h \approx 0.73$.

In supersymmetric (SUSY) models such as the 
minimal supersymmetric standard model
(MSSM)~\cite{CHUNG2005} one typically imposes
a $\dd{Z}_2$
matter parity on the superpotential in order to remove the $B-L$ violating terms
allowed by the SM gauge symmetries. This is equivalent to so called
$R$-parity~\cite{Martin1997},
a $\dd{Z}_2$ symmetry of the Lagrangian under which the charge of
a physical state is related to
its spin. The scalar and fermionic components of a chiral superfield have
opposite $R$-parity and the lightest supersymmetric ($R$-parity odd) particle
(LSP) is stable. In the such models one therefore predicts a new stable particle
that could be a dark matter candidate,
motivated for reasons not \textit{a priori} related to dark matter.
In general the LSP could be either a gravitino or a neutralino, depending for 
example of the nature of SUSY breaking which determines the typical value 
of the gravitino mass.

In such theories a sub-weak-strength interacting neutralino is generally
considered a good candidate for LSP dark matter~\cite{Ellis1984,Jungman1996}.
Since such a particle is
self-charge-conjugate, its relic abundance is determined by thermal
freeze-out, not by matter-antimatter asymmetry as in the case of baryons.
At some point in the early universe the expansion rate of the universe would
have become larger than the LSP's self-annihilation rate
(or co-annihilation rate with other supersymmetric particles). At this point
the number density of LSPs would no longer be able to track its
equilibrium value. It would remain much larger than the equilibrium value,
only diluting due to Hubble expansion. An LSP dark matter candidate with
a larger cross-section would be able to stay in equilibrium longer,
meaning that there would be less of it in the universe today.
Thermal neutralino dark matter has been widely studied in the
MSSM~\cite{Cotta:2009zu,King:2006tf,King:2007vh,Ellis:2007by}
and constrained (c)MSSM~\cite{Kane:1993td,Feng:1999zg,Ellis:1999mm,Ellis:2001nx,Baer:2002fv}.
Successful dark matter may be realised if the LSP is the
lightest neutralino, with various regions of parameter space corresponding to 
different dominant annihilation mechanisms. For example the bulk region corresponds to 
annililation via t-channel slepton exchange, the focus region via t-channel chargino exchange and
the funnel region via s-channel Higgs exchange. There are also other regions
corresponding to co-annihilation with staus or stops.


However the MSSM has certain shortcomings and it is possible that TeV scale SUSY
is realised via a richer structure. It is therefore worthwhile considering alternative SUSY theories
in which dark matter may be realised differently from in the MSSM, with the neutralino dark matter
having a different composition and/or annihilating via different mechanisms.
For example, the $E_6$SSM~\cite{King2006,King:2005my,KING2007}
(or the Exceptional Supersymmetric extension to the Standard Model (SM))
is a string theory inspired supersymmetric model based on an $E_6$
grand unification (GUT) group. The low energy gauge group contains
an extra $U(1)$, called $U(1)_N$, under which
the right-handed neutrinos that arise in the model are not charged.
This means that the right-handed neutrinos may acquire large
intermediate-scale
Majorana masses, leading to a type-I see-saw mechanism to explain
the small observed neutrino masses. This choice, that the low energy gauge group
is $\rr{SM}\times U(1)_N$, defines the model. The $U(1)_N$ gauge symmetry is
spontaneously broken at low energy by a SM-singlet field which we refer to as $S_3$.
This field radiatively acquires a vacuum expectation value (VEV) $s$ naturally of
order the soft SUSY breaking scale, leading to a $Z'$-boson with
an induced mass of order the TeV scale.
Automatic anomaly cancellation is ensured by allowing three complete 27
representations of $E_6$ to survive down to the low energy scale.
These three 27s contain the three generations of known matter, however they also
contain the VEV acquiring Higgs doublets and SM-singlet.
This means that there are two extra copies of the Higgs doublets and SM-singlet
in the low energy particle spectrum. In the \essm only one generation of Higgs doublets and singlets,
defined to be the third, acquires the required VEVs and are called ``active'', 
namely ${H}_{d3}$, ${H}_{u3}$ and $S_3$.
The other two generations, the first and second, of Higgs doublets ${H}_{d\alpha}$, ${H}_{u\alpha}$
and SM-singlets $S_{\alpha}$, where $\alpha \in {1,2}$,
do not acquire VEVs and are called ``inert''.
The inert generations have suppressed Yukawa couplings to SM matter,
suppressing flavour changing neutral currents (FCNCs) and in turn explaining
why they do not radiatively acquire VEVs.

In the MSSM the neutralino mass matrix has the familiar $4 \times 4$
structure corresponding to the bino $\tilde{B}$, the neutral wino 
$\tilde{W}^3$ and two active\footnote{We shall refer to Higgsinos as being
``active'' or ``inert'' according to whether their
scalar SUSY partners do or do not develop VEVs.}
neutral Higgsinos 
$\tilde{H}^0_{d3}$ and $\tilde{H}^0_{u3}$.
In the \essm the neutralino mass matrix is greatly enlarged to a $12 \times 12$ structure
including an additional four neutral inert Higgsinos
$\tilde{H}^0_{d\alpha}$, $\tilde{H}^0_{u\alpha}$, 
one active singlino $\tilde{S}_3$, two inert singlinos $\tilde{S}_{\alpha}$ and
an extra bino $\tilde{B}'$ which is the SUSY partner of the $Z'$.
It has been observed that the six inert Higgsinos and singlinos $\tilde{H}^0_{d\alpha}$, 
$\tilde{H}^0_{u\alpha}$ and $\tilde{S}_{\alpha}$ tend to decouple from the rest of the
neutralino spectrum and it makes sense to consider their $6 \times 6$
mass matrix separately~\cite{Hall2009,Hall2010}.
Moreover the lighest inert neutralino state is predominetly inert singlino in nature and
only acquires mass only via mass mixing with the inert Higgsinos proportional to
the electroweak VEV $v$. It has a suppressed mass of order $v^2/s$,
where $s$ is the SM-singlet VEV~\cite{Hall2009}.
It is therefore natural to suppose that the
LSP arises from the inert neutralino sector and is a superposition of
the inert singlino and inert Higgsino components. It has been shown that it is 
possible to reproduce the observed relic density in this model
by allowing annihilations via an s-channel $Z$-boson, while ensuring that the LSP,
which must be similar in mass to the $Z$, having a mass of about 35--50~GeV,
is not ruled out by LEP constraints on $Z$ decays~\cite{Hall2009}.
It has recently been observed that, in this scenario,
since the LSP always couples rather strongly to the
SM-like Higgs boson, it leads to ``invisible'' Higgs decays at the LHC and 
for the same reason
this dark matter scenario may be discovered or
ruled out by dark matter direct detection experiments in the near
future~\cite{Hall2010}\footnote{Since this paper was submitted, the latest XENON100 direct detection
results have been released~\cite{Aprile:2011hi}. These results severely challenge the previous
dark matter scenario in~\cite{Hall2009,Hall2010} but are consistent with the
bino dark matter scenario as discussed in this paper.}.


In this paper we introduce a new scenario for dark matter in the \essm
in which the dark matter candidate is just the usual bino. 
At first sight having a bino dark matter candidate seems 
impossible since, as already discussed,
the inert singlinos $\tilde{S}_{\alpha}$
naturally have suppressed masses and it is very difficult to make them even as heavy as 
half the $Z$ mass. It is clear that the singlinos will always be lighter than the bino, at least
in the case of gaugino unification, as assumed in this paper. 
To overcome this we propose that the singlinos are exactly massless and decoupled from the 
bino, which is achieved in practice by setting their Yukawa
couplings to zero.
This is easy to do by introducing 
a discrete symmetry $\dd{Z}_2^S$ under which the inert singlinos $\tilde{S}_{\alpha}$
are odd and all other states are even, a scenario we refer to as
the $E_6\dd{Z}_2^S$SSM or \ezssmSTOP
In the \ezssm the inert singlinos $\tilde{S}_{\alpha}$ 
will be denoted as $\tilde{\sigma}$ in order to emphasise their different (massless and decoupled)
nature.
The stable dark matter candidate (DMC)
is then generally mostly bino and the observed dark matter
relic density can be achieved via a novel scenario in which the
bino inelastically scatters off of
SM matter into heavier inert Higgsinos during the time of thermal freeze-out,
keeping the bino in equilibrium long enough to give the desired relic abundance.
Providing the inert Higgsinos are close in mass to the bino this is always possible to arrange,
the only constraint being that the inert Higgsinos satisfy the LEP2 constraint of being heavier than
100~GeV. This in turn implies that the bino must also be heavier than or close to 100~GeV.
These constraints are easy to satisfy and, unlike the inert neutralino LSP dark matter scenario,
we find that successful relic abundance can be achieved within
a GUT-scale-constrained version of the model (the c\ezssm),
assuming a unified soft scalar
mass $m_0$, soft gaugino mass $M_{1/2}$ and soft trilinear mass $A_0$ at
the GUT scale~\cite{Athron2009a,Athron2009,Athron:2011wu}.

It is worth noting that
studies of the constrained (c)\essm~\cite{Athron2009a,Athron2009,Athron:2011wu}
have hitherto neglected to study the
full $12 \times 12$ neutralino mass matrix, only considering the $6 \times 6$ mass matrix of the 
MSSM augmented by the active singlino $\tilde{S}_3$ and the extra bino $\tilde{B}'$,
as in the so called USSM~\cite{Kalinowski:2008iq}. Although the question of dark matter
was addressed in the USSM, the requirement of successful 
relic abundance was not imposed on the c\essm~\cite{Athron2009a,Athron2009,Athron:2011wu}
even though the latter analysis considered the same $6 \times 6$ mass matrix as in the USSM.
Here we shall consider the c\essm with the full $12 \times 12$ neutralino mass matrix,  
including both the USSM and inert neutralinos, under the
assumption that the fermionic components of the inert SM-singlet superfields,
the two inert singlinos, are forbidden to acquire mass by an extra $\dd{Z}_2$
symmetry of the superpotential. In practice this reduces to a $10 \times 10$
neutralino mass matrix once the two massless inert singlinos are decoupled.

In summary, the main result of this paper is
that bino dark matter, with nearby inert Higgsinos and massless inert singlinos,  
provides a simple and successful picture of dark matter in the \essm consistent with 
GUT constrained soft parameters.
We shall also consider the effect of 
the presence of the two massless inert singlinos in the \ezssm on the 
effective number of neutrinos contributing to the expansion rate of the universe
prior to BBN, affecting $^4$He production. Current fits to WMAP data~\cite{Komatsu2011}
favour values greater than three, so the presence of additional contributions to the effective number
of neutrinos is another interesting aspect of the  \ezssm which we shall study.
In practice we find that the additional number of effective neutrino species is 
less than two, due to entropy dilution, depending on the mass of the $Z'$ which keeps the
inert singlinos in equilibrium. 


The \ezssm is introduced in Section~\ref{model}.
Section~\ref{neutralinos} explores the neutralino and chargino sectors of the
\ezssm whereas the details of the dark matter calculation are
presented in Section~\ref{dm}.
$\neff$ is defined and calculations of its value in the \ezssm are presented
in Section~\ref{neff}. Some benchmark points are presented in
Section~\ref{benchmarks} and the conclusions are summarised in
Section~\ref{conclusions}.

\section{The \ezssm}
\label{model}
The $E_6$ GUT group can be broken as follows:
\be
E_6 &\rightarrow& SO(10) \times U(1)_\psi \\
&\rightarrow& SU(5) \times U(1)_\chi \times U(1)_\psi \\
&\rightarrow& SU(3) \times SU(2) \times U(1)_Y
\times U(1)_\chi \times U(1)_\psi.
\ee
In the \essm $E_6$ is broken at the GUT scale directly to
$SU(3) \times SU(2) \times U(1)_Y \times U(1)_N$, where
\be
U(1)_N &=& \cos(\vartheta) U(1)_\chi + \sin(\vartheta) U(1)_\psi
\ee
and $\tan(\vartheta) = \sqrt{15}$ such that the right-handed neutrinos
are chargeless. Three complete 27 representations of $E_6$ then survive
down to low energy in order to ensure anomaly cancellation. These $27_i$ ($i \in
\{1,2,3\}$) decompose under the $SU(5) \times U(1)_N$ subgroup as follows:
\be
27_i &\rightarrow& \Bigl(10,1/\sqrt{40}\Bigr)_i + \Bigl(\bar{5},2/\sqrt{40}\Bigr)_i
\nn\\
&& +\ \Bigl(\bar{5},-3/\sqrt{40}\Bigr)_i + \Bigl(5,-2/\sqrt{40}\Bigr)_i
+ \Bigl(1,5/\sqrt{40}\Bigr)_i + \Bigl(1,0\Bigr)_i
\ee
The first two terms contain normal matter, whereas the final term, which is
a singlet under the entire low energy gauge group, contains the right-handed
neutrinos. The second-to-last term, which is charged only under $U(1)_N$,
contains
the SM-singlet fields $S_i$. The third generation SM-singlet acquires
a VEV $\langle S_3\rangle = s/\sqrt{2}$ which, as we shall see,
generates the effective $\mu$-term and spontaneously breaks
$U(1)_N$ leading to a mass for the
$Z'$-boson.
The remaining terms contain three generations of down- and up-type
Higgs doublets $H_{di}$ and $H_{ui}$ as well as exotic coloured states
$\bar{D}_i$ and $D_i$. Again only the third generation of Higgs doublets
acquire VEVs $\langle H^0_{d3}\rangle = v_d/\sqrt{2} = v\cos(\beta)/\sqrt{2}$ and
$\langle H^0_{u3}\rangle = v_u/\sqrt{2} = v\sin(\beta)/\sqrt{2}$.

The low energy gauge invariant superpotential can be written as follows:
\be
\cc{W} &=& \cc{W}_0 + \cc{W}_1 + \cc{W}_2,
\label{w}\ee
where
\be
\cc{W}_0 &=& \lambda_{ijk}S_i H_{dj} H_{uk} + \kappa_{ijk} S_i D_j \bar{D}_k
+ h^N_{ijk} N^c_i H_{uj} L_k \nn\\
&& +\ h^U_{ijk} u^c_i H_{uj} Q_k + h^D_{ijk} d^c_i H_{dj} Q_k
+ h^E_{ijk} e^c_i H_{dj} L_k, \\
\cc{W}_1 &=& g^Q_{ijk} D_i Q_j Q_k + g^q_{ijk} \bar{D}_i d^c_j u^c_k, \\
\cc{W}_2 &=& g^N_{ijk} N^c_i D_j d^c_k + g^E_{ijk} e^c_i D_j u^c_k
+ g^D_{ijk} Q_i L_j \bar{D}_k.
\ee
It is now clear that the effective $\mu$-parameter is given by
$\mu = \lambda_{333}s/\sqrt{2}$ generating the term $\mu H_{d3} H_{u3}$ in the
superpotential.

In order to suppress non-diagonal flavour transitions arising
from the Higgs sector the superpotential obeys an approximate $\dd{Z}_2$
symmetry called $\dd{Z}_2^H$. Under this symmetry all superfields other
than $S_3$, $H_{d3}$ and $H_{u3}$ are odd. It is this approximate symmetry
that distinguishes between the third generation and the inert generations of
Higgs doublets and SM-singlets, with the inert generations having suppressed
couplings to matter and not radiatively acquiring VEVs.
This approximate symmetry suppresses $\lambda_{ijk}$ couplings of
the forms $\lambda_{\alpha 33}$, $\lambda_{3\alpha 3}$, $\lambda_{33\alpha}$
and $\lambda_{\alpha\beta\gamma}$, where $\alpha,\beta,\gamma \in \{1,2\}$
\textit{i.e.} labelling the inert generations only.
Such an approximate
$\dd{Z}_2^H$ symmetry, with a stable hierarchy of couplings, can
be realised in \essm flavour theories such as the one proposed by Howl
\textit{et al.}~\cite{Howl2010}.
The symmetry cannot be exact or else the lightest of the exotic coloured
states would be absolutely stable which would contradict observation.

Given this, an exact $\dd{Z}_2$ symmetry must be imposed on the superpotential
in order to avoid rapid proton decay.
There are two ways to impose an appropriate $\dd{Z}_2$ symmetry on
$\cc{W}$ that leads to baryon and lepton number conservation.
The first option is to impose a symmetry called $\dd{Z}_2^L$ under which
only the lepton superfields are odd.
In this case the superpotential is equal to $\cc{W}_0 + \cc{W}_1$ and
the model is called the $E_6$SSM-I.
$U(1)_B$ and $U(1)_L$ are symmetries of the superpotential
if the exotic coloured states $\bar{D}$ and $D$ are interpreted as
diquarks and antidiquarks, with ($B = \pm2$, $L = 0$).
The second option is to impose a symmetry called $\dd{Z}_2^B$ under which
both the lepton superfields and the
exotic $\bar{D}$ and $D$ superfields are odd.
In this case the superpotential is equal to $\cc{W}_0 + \cc{W}_2$ and
the model is called the $E_6$SSM-II.
$U(1)_B$ and $U(1)_L$ are symmetries of the superpotential
if the exotic coloured states are interpreted as leptoquarks, with
($B = \mp1$, $L = \mp1$).

It needs to be noted that the superpotential of the \essm is already
automatically invariant
under the usual matter parity of the MSSM, provided that the
exotic $\bar{D}$ and $D$ superfields as well as the Higgs SM-singlet superfields
are interpreted as being even under matter parity, along with the Higgs doublets.
The $B - L$ violating terms of the MSSM superpotential that
matter parity is invoked to forbid are never present in the \essm superpotential
of Eq.~(\ref{w}) since they would
violate the extra surviving $U(1)_N$ gauge symmetry contained in $E_6$.
We shall refer to this usual matter parity of the MSSM which automatically
arises in this model as $\dd{Z}_2^M$. In the usual way it can be recast as
$R$-parity in terms of the superfield components, with the scalar components 
of the superfield being assigned the same parity as the superfield, 
and the fermionic components being assigned the opposite parity.
As in the MSSM, the states which are odd under $R$-parity
are called the superpartners, 
with the lightest superpartner being absolutely stable.

In the \ezssm the superpotential is also invariant under an additional exact
$\dd{Z}_2$ symmetry called $\dd{Z}_2^S$. Under this symmetry only
the two inert SM-singlet superfields $S_\alpha$ are odd.
The couplings of the forms
$\lambda_{\alpha ij}$ and $\kappa_{\alpha ij}$ are forbidden. This
means that the fermionic components of $S_\alpha$, the inert singlinos
$\tilde{\sigma}$, are
forbidden to have mass and do not mix with the other neutralinos.
They only interact via their gauge couplings to the $Z'$-boson, which
exist since they
are charged under the extra $U(1)_N$ gauge symmetry.
The extra $\dd{Z}_2^S$ symmetry of the superpotential does not change the forms of
the mass matrices of the
the non-inert sectors of the model. In particular this means that all of
the squark, slepton, gluino and non-inert Higgs scalar masses and mass matrices
are the same as the
ones given in \cite{Athron2009}.
The issue of $Z$-$Z'$ mixing is also the same as
discussed there. 

All of the exact and approximate $\dd{Z}_2$ symmetries of the superpotential
mentioned in this section are summarised in Table~\ref{z2s}.

\begin{table}[h]\begin{center}
\begin{tabular}{r|ccccc|}
& $\dd{Z}_2^H$ & $\dd{Z}_2^L$ & $\dd{Z}_2^B$ & $\dd{Z}_2^M$ & $\dd{Z}_2^S$ \\\hline
$S_\alpha$                & $-$ & $+$ & $+$ & $+$ & $-$ \\
$H_{d\alpha},H_{u\alpha}$ & $-$ & $+$ & $+$ & $+$ & $+$ \\
$S_3$                     & $+$ & $+$ & $+$ & $+$ & $+$ \\
$H_{d3},H_{u3}$           & $+$ & $+$ & $+$ & $+$ & $+$ \\
$Q_i,u^c_i,d^c_i$         & $-$ & $+$ & $+$ & $-$ & $+$ \\
$L_i,e^c_i$               & $-$ & $-$ & $-$ & $-$ & $+$ \\
$\bar{D}_i,D_i$           & $-$ & $+$ & $-$ & $+$ & $+$ \\\hline
\end{tabular}
\caption{The transformations of the superfields under the
various $\dd{Z}_2$ symmetries of the superpotential
that are mentioned in this paper.
$\dd{Z}_2^H$ is an approximate flavour symmetry.
Either $\dd{Z}_2^L$ or $\dd{Z}_2^B$ is imposed in order to avoid rapid proton decay.
$\dd{Z}_2^M$ matter parity is already a symmetry of the \essm due to gauge
symmetry. $\dd{Z}_2^S$ is the extra symmetry which is 
imposed in the EZSSM, forcing the inert singlinos
to be massless.\label{z2s}}
\end{center}\end{table}

It is known that the model as presented thus far leads
to gauge coupling unification at too high a scale, with the GUT scale
typically being higher than the Planck scale. This issue can be solved by
having the $E_6$ GUT group be broken to an intermediate group before being
broken finally to $\rr{SM} \times U(1)_N$ as shown in
\cite{Howl2008}. In this paper, however, for simplicity,
we implement the
usual solution where the superpotential contains a bilinear term
involving extra fields from incomplete $27'$ and $\overline{27'}$
representations $\cc{W}' = \mu' H' \overline{H'}$.
To some extent this solution reintroduces the
$\mu$ problem, but $\mu'$ is not required to be related to the electroweak
symmetry breaking (EWSB) scale and in order to observe satisfactory
gauge coupling unification it is only required that $\mu' \lesssim 100$~TeV.

\section{The Neutralinos and Charginos of the \ezssm}
\label{neutralinos}
In the EZSSM the chargino mass matrix in the interaction basis
\be
\tilde{C}_{\rr{int}} &=& \left(\ba{c} \tilde{C}^+_{\rr{int}} \\
\tilde{C}^-_{\rr{int}} \ea\right),
\ee
where
\be
\tilde{C}^+_{\rr{int}} = \left(\ba{c}
\tilde{W}^+ \\ \tilde{H}_{u3}^+ \\\hline \tilde{H}_{u2}^+ \\ \tilde{H}_{u1}^+
\ea\right) &\rr{and}& 
\tilde{C}^-_{\rr{int}} = \left(\ba{c}
\tilde{W}^- \\ \tilde{H}_{d3}^- \\\hline \tilde{H}_{d2}^- \\ \tilde{H}_{d1}^-
\ea\right),
\ee
is given by
\be
M^C &=& \left(\ba{cc} & P^T \\ P \ea\right),
\ee
where
\be
P = \left(\ba{cc|cc} M_2 & \sqrt{2}m_Ws_\beta & & \\
\sqrt{2}m_Wc_\beta & \mu & \frac{1}{\sqrt{2}}\lambda_{332}s & \frac{1}{\sqrt{2}}\lambda_{331}s \\\hline
& \frac{1}{\sqrt{2}}\lambda_{323}s & \frac{1}{\sqrt{2}}\lambda_{322}s & \frac{1}{\sqrt{2}}\lambda_{321}s \\
& \frac{1}{\sqrt{2}}\lambda_{313}s & \frac{1}{\sqrt{2}}\lambda_{312}s & \frac{1}{\sqrt{2}}\lambda_{311}s
\ea\right).
\ee
The top-left block is the MSSM chargino mass matrix whereas the bottom-right
block contains mass terms for the extra inert Higgsino states.
The Yukawa couplings in the off-diagonal blocks are suppressed under
the approximate $\dd{Z}_2^H$ and
are therefore expected to be small, implying that there
should not be much mixing between the MSSM and inert states.

We define the term ``neutralino'' not to include the massless
inert singlinos
which have no Yukawa couplings involving them and are decoupled.
The neutralino mass matrix $M^N$ in the interaction basis
\be
\tilde{N}_{\rr{int}} &=& \left(\ba{cccccc|cc}
\tilde{B} & \tilde{W}^3 & \tilde{H}^0_{d3} & \tilde{H}^0_{u3} & \tilde{S}_3 &
\tilde{B}' & \tilde{H}^0_{d\alpha} & \tilde{H}^0_{u\beta} \ea\right)^T
\ee
is then equal to
\be
\left(\ba{cccccc|cc}
M_1 & 0 & -\frac{1}{2}g'v_d & \frac{1}{2}g'v_u & & & & \\
0 & M_2 & \frac{1}{2}gv_d & -\frac{1}{2}gv_u & & & & \\
-\frac{1}{2}g'v_d & \frac{1}{2}gv_d & 0 & -\mu & -\frac{\lambda_{333}v_u}{\sqrt{2}} & Q_dg_1'v_d &
0 & -\frac{\lambda_{33\beta}s}{\sqrt{2}} \\
\frac{1}{2}g'v_u & -\frac{1}{2}gv_u & -\mu & 0 & -\frac{\lambda_{333}v_d}{\sqrt{2}} & Q_ug_1'v_u &
-\frac{\lambda_{3\alpha 3}s}{\sqrt{2}} & 0 \\
& & -\frac{\lambda_{333}v_u}{\sqrt{2}} & -\frac{\lambda_{333}v_d}{\sqrt{2}} & 0 & Q_Sg_1's &
-\frac{\lambda_{3\alpha 3}v_u}{\sqrt{2}} & -\frac{\lambda_{33\beta}v_d}{\sqrt{2}} \\
& & Q_dg_1'v_d & Q_ug_1'v_u & Q_Sg_1's & M_1' & & \\\hline
& & 0 & -\frac{\lambda_{3\alpha 3}s}{\sqrt{2}} & -\frac{\lambda_{3\alpha 3}v_u}{\sqrt{2}} & &
0 & -\frac{\lambda_{3\alpha\beta}s}{\sqrt{2}} \\
& & -\frac{\lambda_{33\beta}s}{\sqrt{2}} & 0 & -\frac{\lambda_{33\beta}v_d}{\sqrt{2}} & &
-\frac{\lambda_{3\alpha\beta}s}{\sqrt{2}} & 0
\ea\right), \label{nmtx}
\ee
where once again $\alpha,\beta \in \{1,2\}$, indexing the inert generations.
$Q_d = -\frac{3}{\sqrt{40}}$, $Q_u = -\frac{2}{\sqrt{40}}$ and $Q_S = \frac{5}{\sqrt{40}}$ are the $U(1)_N$
charges of down-type Higgs doublets, up-type Higgs doublets and
SM-singlets respectively and $g_1'$ is the GUT normalised $U(1)_N$
gauge coupling.
$M_1$, $M_2$ and $M_1'$ are soft gaugino masses. Typically $g_1' \approx g_1$
all the way down to the low energy scale. If the soft gaugino masses are unified
at the GUT scale (universal $M_{1/2}$) then we also have $M_1' \approx M_1
\approx M_2/2$. The
matrix as written neglects the small kinetic term mixing between $\tilde{B}$ and
$\tilde{B}'$. The elements left empty in this matrix and similar ones
are implicitly taken to be zero.

The Yukawa couplings in the off-diagonal blocks are suppressed under the
approximate $\dd{Z}_2^H$. Given the smallness of these couplings, the
inert neutralinos in the bottom-right block are pseudo-Dirac states with
an approximately decoupled mass matrix
\be
-\frac{s}{\sqrt{2}}\left(\ba{cccc}
& & \lambda_{322} & \lambda_{321} \\
& & \lambda_{312} & \lambda_{311} \\
\lambda_{322} & \lambda_{312} & & \\
\lambda_{321} & \lambda_{311} & &
\ea\right)
& \mbox{in the basis} & \left(\ba{cccc}
\tilde{H}^0_{d2} & \tilde{H}^0_{d1} & \tilde{H}^0_{u2} & \tilde{H}^0_{u1}
\ea\right). \nn
\ee
They are approximately degenerate with the two inert chargino Dirac states.

The top-left block contains the states of the MSSM supplemented by the
third generation singlino and the bino$'$. This is known as the
USSM sector~\cite{Kalinowski2009}. In the case where $M_1 \approx
M_1'$ is small the lightest neutralino mass state will be mostly bino. The
bino$'$ will mix with the third generation singlino giving two mixed states
with masses around $Q_Sg_1's$.
As $M_1 \approx M_1'$ increases the bino mass will increase relative to both the
third generation Higgsino mass $\mu$ and the inert Higgsino masses given
approximately by the bi-unitary diagonalisation of
$-\frac{1}{\sqrt{2}}\lambda_{3\alpha\beta}s$. At the same time the state mostly
containing the third generation singlino will have a decreasing mass as $M_1'$
increases relative to $Q_Sg_1's$.

\section{Dark Matter in the c\ezssm}
\label{dm}
As discussed previously, due to the automatic matter parity of the model,
there is a conserved $R$-parity under which
the charginos, neutralinos, inert singlinos $\tilde{\sigma}$ and
exotic $\bar{D}$ and $D$ fermions,
along with the squarks and sleptons, are all $R$-parity odd
\textit{i.e.} all of the fermions
other than the quarks and leptons are $R$-parity odd.
We shall assume that the
lightest neutralino $\tilde{N}_1$ is the lightest of all of the
$R$-parity odd states,
excluding the massless inert singlinos $\tilde{\sigma}$.
However $\tilde{N}_1$ cannot decay into $\tilde{\sigma}$ via neutralino mixing
since the inert singlinos are decoupled from the neutralino mass matrix.
Furthermore the possible
decay $\tilde{N}_1 \rightarrow \tilde{\sigma}\sigma$,
allowed by the $\sigma$-$\tilde{\sigma}$-$\tilde{B}'$ supersymmetric $U(1)_N$
gauge coupling,
is forbidden if $\tilde{N}_1$ is lighter than the inert SM-singlet scalars $\sigma$.
In fact, in this case, no kinematically viable final states exist that have
the same quantum
numbers as $\tilde{N}_1$. Therefore $\tilde{N}_1$ is absolutely
stable and is the DMC of the model.

In the successful dark matter scenario presented in this section $\tilde{N}_1$
has a dominant bino $\tilde{B}$ component with at least one of the
two pairs pseudo-Dirac inert Higgsinos
expected to be close in mass, but somewhat heavier,
in order to achieve the correct relic density.
This is due to a novel scenario in which 
the DMC, approximately the bino, inelastically scatters off of
SM matter into heavier inert Higgsinos during the time of thermal freeze-out,
keeping it in equilibrium long enough to give a successful relic density. 
In this section we discuss in detail how this novel scenario
comes about in this model.

\subsection{The Dark Matter Calculation}
Usually in supersymmetric models the evolution of the cosmological
number density $n_i$ of a supersymmetric
($R$-parity odd) particle $i$ in the early universe can be expressed as
\be
\dot{n}_i &=& -3Hn_i - \sum_j \langle \sigma_{ij}v_{ij}\rangle \bigl(n_in_j - n_i^\eq n_j^\eq\bigr) \nn\\
&& -\sum_{j\ne i} \Bigl[\Gamma_{ij}\bigl(n_i-n_i^\eq\bigr) - \Gamma_{ji}\bigl(n_j - n_j^\eq\bigr)\Bigr] \nn\\
&& -\sum_{j\ne i}\sum_X \Bigl[\langle\sigma'_{Xij}v_{iX}\rangle\bigl(n_in_X - n_i^\eq n_X^\eq\bigr)
- \langle\sigma'_{Xji}v_{jX}\rangle\bigl(n_jn_X - n_j^\eq n_X^\eq\bigr)\Bigr].
\label{dotni}\ee
The first term accounts for Hubble expansion and the second term accounts for
annihilations with other supersymmetric particles,
including self-annihilations. The third term represents the decays
of supersymmetric particles $i$ into other supersymmetric species $j$
as well decays of other supersymmetric species into species $i$. The final
term represents the inelastic scattering of supersymmetric particles $i$ off of
$R$-parity even particles $X$ into other supersymmetric species $j$
and \textit{vice versa}~\cite{Griest:1990kh,Schelke2004}.

Summing up these equations yields the somewhat simpler
\be
\dot{n} \equiv \sum_i\dot{n}_i &=&
-3Hn - \sum_i\sum_j\langle\sigma_{ij}v_{ij}\rangle\bigl(n_in_j - n_i^\eq n_j^\eq\bigr). \label{nsum}
\label{dotn}\ee
It should be noted that, assuming all supersymmetric particles can decay into
the DMC in a reasonable amount of time, after thermal freeze-out the
relic number density of the DMC will subsequently becomes equal to $n$.

In our model the DMC
is not the lightest $R$-parity odd state (an inert singlino), but
the lightest neutralino $\tilde{N}_1$.
We would like to use Eq.~(\ref{dotni}) to describe the evolution of $R$-parity
odd states other than the inert singlinos, generically $\tilde{\chi}$.
In this case we should also include
in Eq.~(\ref{dotni}) processes involving $\sigma$ and $\tilde{\sigma}$
particles that change the number of $\tilde{\chi}$ particles by one.
Since such processes necessarily
involve inert SM-singlet scalars $\sigma$,
it is valid to neglect these processes
in the case where these inert SM-singlets have frozen out long before the
freeze-out of dark matter. We will call this condition~1 and it should be
satisfied
given our assumption that the inert SM-singlet scalars are heavier than the DMC,
since they only interact via the heavy $Z'$-boson.
As we shall see, the value of $n$ after the thermal freeze-out of $\tilde{N}_1$
depends on
annihilation cross-sections involving $\tilde{N}_1$ and other $R$-parity odd
states close by in mass. As long as condition~1 is satisfied, meaning that
we can neglect such annihilations that also have
inert singlinos in the final state during thermal freeze-out,
we can neglect the inert singlinos
and use Eq.~(\ref{dotn}) to calculate the number density of $R$-parity states
other than inert singlinos. $n$ will eventually be equal to the number density
of DMCs after other $\tilde{\chi}$ particles have decayed to $\tilde{N}_1$.

During thermal freeze-out the annihilation rates of
the $\tilde{\chi}$ particles become small compared to
the expansion rate of the universe and their number densities
become larger than their (non-relativistic) equilibrium values.
The universe expands too
fast for the number densities to track their equilibrium values.
Let us assume however that these states inelastically scatter off of
SM states $X$
frequently enough that the ratios of the number densities of the $\tilde{\chi}$
particles do maintain their equilibrium values during the time of
thermal freeze-out.
We shall call this condition~2 and assuming that it is satisfied we have
\be
\frac{n_j}{n_i} = \frac{n_j^\eq}{n_i^\eq} &\Rightarrow&
\frac{n_i}{n} = \frac{n_i^\eq}{n_\eq},
\ee
which allows us to rewrite Eq.~(\ref{nsum}) as
\be
\dot{n} = -3Hn - \langle\sigma v\rangle\bigl(n^2 - n_\eq^2\bigr),
\ee
where $n_\eq = \sum_in_i^\eq$ and
\be
\langle\sigma v\rangle &=& \sum_i\sum_j\langle\sigma_{ij}v_{ij}\rangle
\frac{n_i^\eq n_j^\eq}{n_\eq^2}.
\ee
Here we see that since heavier $\tilde{\chi}$ particles would have smaller
non-relativistic
equilibrium number densities there would be fewer of them around during
the dark matter's thermal freeze-out and
annihilation cross-sections involving them would be less important.

\subsection{The c\ezssm}
In order to carry out the dark matter analysis in the constrained version of
the model we have extended the RGE code used by
Athron \textit{et al.}~\cite{Athron2009} to include
the Yukawa parameters and soft masses of the inert sector of the
\ezssmSTOP The inputs are $\kappa_{3ij}$ and $\lambda_{333}$ at the GUT scale,
$\lambda_{3\alpha\beta}$ at the EWSB scale, $s$ and $\tan(\beta)$,
as well as the known low energy Yukawa couplings and gauge couplings. Given
these inputs and the RGEs the algorithm attempts to find points
with GUT scale unified soft masses $m_0$, $M_{1/2}$ and $A_0$. The low energy
$U(1)_N$ gauge coupling $g_1'$ is set by requiring it to be equal to
the other gauge couplings at the GUT scale.

For consistent points in the \essm
the lightest non-inert (USSM sector)
supersymmetric particle is typically bino dominated.
For the c\ezssm we find the same thing. The masses of the inert Higgsino states
depend on $s$ and on the the Yukawa couplings $\lambda_{3\alpha\beta}$ and
the in the c\ezssm the lightest neutralino can be either the bino dominated
state or a pseudo-Dirac inert Higgsino dominated state. In the latter case
we find that the DMC pseudo-Dirac inert Higgsino states co-annihilate with
full-weak-strength interactions and lead to a too small
dark matter relic density.
In the former case the bino DMC normally annihilates too weakly and yields
a too large dark matter relic density. If, however, there are inert Higgsino
states close by in mass, they contribute significantly to
$\langle\sigma v\rangle$ allowing for the observed amount of dark matter.
This relies on condition~2 being satisfied \textit{i.e.} the binos being
up-scattered into inert Higgsinos with a large enough rate.

Such points with consistent dark matter relic density can be found and three
are presented in Section~\ref{benchmarks}. Condition~1 is satisfied since
the inert SM-singlet scalars are so much heavier than the DMC and
the $Z'$-boson mass is so large compared to the regular $Z$-boson mass.
Annihilation and scattering processes involving inert SM-singlets and singlinos
must contain a virtual $Z'$-boson.

To test condition~2 let us compare the rate for binos up-scattering into
inert Higgsinos with the inert Higgsino co-annihilation rate. We shall label
the mostly bino state $\tilde{N}_1$ and the lightest pseudo-Dirac inert Higgsino
states $\tilde{N}_2$ and $\tilde{N}_3$. The dominant up-scattering diagrams are
of the following form:
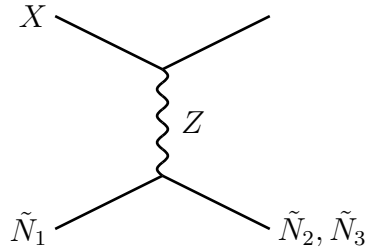
\begin{center}
\begin{picture}(120,120)(-60,-60)
\SetWidth{1}
\Line(-40,40)(0,20)
\Line(0,20)(40,40)
\Photon(0,20)(0,-20){2}{4}
\Line(-40,-40)(0,-20)
\Line(0,-20)(40,-40)
\Text(-43,40)[r]{$X$}
\Text(-43,-40)[r]{$\tilde{N}_1$}
\Text(43,-40)[l]{$\tilde{N}_{2},\tilde{N}_{3}$}
\Text(7,0)[l]{$Z$}
\end{picture}
\end{center}
We define $R_{Zij}$ couplings such that the
$Z$-$\tilde{N}_i$-$\tilde{N}_j$ coupling is equal to $R_{Zij}$ times
the $Z$-$\nu$-$\nu$ coupling. We have
\be
R_{Zij} &=& \sum_{D=3,7,9}N_i^DN_j^D - \sum_{U=4,8,10}N_i^UN_j^U,
\ee
where $N_i^a$ is the neutralino mixing matrix element corresponding to
mass eigenstate $i$ and interaction state $a$. $D$ and $U$ index the
down- and up-type Higgsino interaction states respectively. For the
pseudo-Dirac inert Higgsino states we have
\be
m_3 \approx -m_2 &\mbox{and}& R_{Z23} \approx 1,
\ee
allowing for full-weak-strength co-annihilations of the following form:
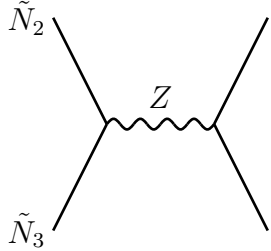
\begin{center}
\begin{picture}(120,120)(-60,-60)
\SetWidth{1}
\Line(-40,40)(-20,0)
\Line(-20,0)(-40,-40)
\Photon(-20,0)(20,0){2}{4}
\Line(40,40)(20,0)
\Line(20,0)(40,-40)
\Text(-43,40)[r]{$\tilde{N}_2$}
\Text(-43,-40)[r]{$\tilde{N}_3$}
\Text(0,10)[c]{$Z$}
\end{picture}
\end{center}

The ratio of the rate for the mostly bino state up-scattering into
the mostly inert Higgsino state to the inert Higgsino co-annihilation rate
is given approximately by
\be
\Upsilon &=& \frac{\langle\sigma'_{X12}v_{1X}\rangle n_1^\eq n_X^\eq}
{\langle\sigma_{23}v_{23}\rangle n_2^\eq n_3^\eq}
\ee
To give an idea of the size of this ratio, if the SM particle $X$ is
relativistic and $m_1 \sim m_2 \approx m_3$ then
\be
\Upsilon &\sim& \left(\frac{R_{Z12}}{R_{Z23}}\right)^2 \frac{T^3}{(|m_1|T)^{3/2}\exp(-|m_1|/T)} \nn\\
&\approx& R_{Z12}^2\left(\frac{1}{x}\right)^{3/2}e^x,
\ee
where $x = |m_1|/T$ and $T$ is the temperature.
This ratio is expected to be large because of the overwhelming abundance
of the relativistic SM particle $X$, but it also depends on $R_{Z12}$.
The value of $R_{Z12}$ depends
on the $\dd{Z}_2^H$-breaking couplings that mix the top-left block of the
neutralino mass matrix in Eq.~(\ref{nmtx}), the USSM states
including the bino, with the
inert Higgsino states in the bottom-right block. Since this symmetry is
not exact we expect these couplings to be large enough such that we can
still assume $\Upsilon \gg 1$. Explicit examples of this parameter
are included in Table~\ref{dmtable} in Section~\ref{benchmarks}.

With the two conditions satisfied we use \texttt{micrOMEGAs}~\cite{Belanger2010}
to calculate
the dark matter relic density for low energy spectra consistent with the
GUT-scale-constrained scenario. The \texttt{CalcHEP} model files for the \ezssm
are produced using \texttt{LanHEP}~\cite{Semenov2010}.
The observed relic density of dark matter can
arise in this model and examples are shown in Section~\ref{benchmarks}. The most
critical factor is the mass splitting between the bino and the lightest inert
Higgsinos. Too large and there would not be enough inert Higgsinos remaining at
the time of the bino's thermal freeze-out to have a significant enough effect.
Too small and $\langle\sigma v\rangle$ would be dominated by inert Higgsino
co-annihilations leading to a too small dark matter relic density.

Since in this scenario the DMC is predominantly bino, the spin-independent
DMC-nucleon cross-section $\sigma_{\rr{SI}}$ is not expected to be in the range
that
direct detection experiments are currently sensitive too. The spin-independent
cross-section of a pure bino is suppressed by the squark masses,
but is also sensitive
to the squark mixing angles~\cite{Choi:2000kh}. For each flavour
the cross-section vanishes for zero squark mixing.
Since in practice the DMC will also have non-zero (but small)
active Higgsino components, there are also contributions to $\sigma_{\rr{SI}}$
from t-channel active Higgs scalar exchange, via the bino-Higgs-Higgsino
supersymmetric gauge coupling. These contributions, though dominant,
are quite small
due to the overwhelming bino nature of the DMC. Estimates of
$\sigma_{\rr{SI}}$, using the same proton $f_d$, $f_u$ and $f_s$ parameters
used by Gogoladze \textit{et al.}~\cite{Gogoladze:2010ch},
are included in Table~\ref{dmtable}
in Section~\ref{benchmarks}.

\section{The Inert Singlinos and Their Contribution to the Effective Number of
Neutrinos prior to BBN}
\label{neff}

In the standard theory of BBN, which happens long after
the thermal freeze-out of dark matter,
the resultant primordial abundances of
the light elements depend on two parameters---the effective number of
neutrinos contributing to the expansion rate of the radiation dominated
universe $\neff$ and the baryon-to-photon ratio $\eta$.

Whilst the
primordial abundance of $^4$He is not the most sensitive measure of $\eta$,
it is much more sensitive to $\neff$ than the other light element abundances.
This is because prior to nucleosynthesis when the equilibrium photon temperature
is of order
0.1~MeV the number of neutrons remaining, virtually all of which are
subsequently incorporated into $^4$He nuclei, is sensitive to the expansion rate
of the universe, which depends on $\neff$. The greater the expansion rate, the
less time there is for charged current weak interactions to convert neutrons
into protons.

The analysis by Izotov \textit{et al.}~\cite{Izotov2010}
using the more recent neutron lifetime measurement
by Serebrov \textit{et al.}~\cite{Serebrov2008}
gives $\neff = 3.80^{+0.80}_{-0.70}$ at 2-sigma, implying a more-than-2-sigma
tension between the measured $^4$He abundance and the Standard Model prediction
for $\neff$ (about 3). Although Aver \textit{et al.}~\cite{Aver2010} suggest
that these errors may be larger,
similar results are also obtained for the effective number of neutrinos
contributing to the expansion rate of the
universe from fits to WMAP data~\cite{Komatsu2011}.

In the \ezssm the two massless inert singlinos would have decoupled from
equilibrium at an earlier time than the light neutrinos, but nevertheless
would have contributed to $\neff$. Exactly when the inert singlinos would have
decoupled from equilibrium with the photon depends on
the mass of the $Z'$-boson which determines the strength of an effective
Fermi-like 4-point interaction vertex that would have been responsible for
keeping the inert singlinos in equilibrium. The various values for $\neff$
that can be achieved in this model all fit the data better than the SM value.

The implications of extra neutrino-like particles present in the early
universe have long been studied and the methods used
in following analysis rely on relatively simple physics~\cite{Steigman1979}.
The cosmological energy density of a
relativistic $\rr{boson}, \rr{fermion}$ $i$ with
number of degrees of freedom $g^i$ and temperature $T^i$ is given by
\be
\rho^i &=& (1,\nicefrac{7}{8})g^i\frac{\pi^2}{30}(T^i)^4.
\ee
The total radiation dominated energy density is then defined to be
\be
\rho &=& g^\eff\frac{\pi^2}{30}T^4,
\ee
where $T = T^\gamma$ is the photon temperature.
The effective number of degrees of freedom $g^\eff$
takes into account the factor of $\nicefrac{7}{8}$ for fermions and also takes
into account the fact that some species no longer in equilibrium
with the photon may have a different temperature.
In this radiation dominated universe the expansion rate is then given by
\be
H^2 &=& \frac{8\pi G}{3}\rho \\
&=& \frac{1}{M_{\rr{Planck}}^2}g^\eff\frac{4\pi^3}{45}T^4 \\
&\equiv& k_1^2g^\eff T^4,
\ee
where we define the constant $k_1$ for future convenience.

The effective number of degrees of freedom contributing to the expansion rate
of the universe during the run-up to nucleosynthesis is defined to be
\be
g^\eff_0 &=& g^\gamma + \nicefrac{7}{8}g^\nu N_\eff(\nicefrac{4}{11})^{4/3}. \label{411}\\
&=& 2 + \nicefrac{7}{4}N_\eff(\nicefrac{4}{11})^{4/3}
\ee
Here $g^\gamma = 2$ is the number of degrees of freedom of the photon and
$g^\nu = 2$ is the number of degrees of freedom of a light neutrino.
The three SM neutrinos are expected to decouple from equilibrium with
the photon at a temperature above the electron mass whereas nucleosynthesis
does not happen until the temperature is below the electron mass. When the
photon/electron temperature is around the electron mass the electrons
and positrons
effectively disappear from the universe\footnote{A much smaller number of electrons
remains due to the small lepton number asymmetry.}.
Their disappearance heats the photons
to a higher temperature then they would otherwise have had, but the neutrinos,
having already decoupled, would continue to cool at the full rate dictated
by Hubble expansion. Because of the neutrinos' lower temperature
at nucleosynthesis they would contribute less to $g^\eff_0$ per degree of freedom.
Eq.~(\ref{411}) is defined such that in the SM $N_\eff = 3$, for the three
neutrinos decoupling above the electron mass (as we shall see). Extra
particles, such as the \ezssm inert singlinos, decoupling above the muon
mass would have even lower temperatures at the time of nucleosynthesis and
would therefore contribute to $g^\eff_0$ even less than light neutrinos
per degree of freedom.

\subsection{The Calculation of $N_\eff$}
In the c\ezssm there is a typical scenario in which the massless inert singlinos
$\tilde\sigma$ decouple at a temperature above the colour transition temperature
(when the effective degrees of freedom are quarks and gluons
rather than mesons) and above the strange quark mass,
but below the charm quark mass. This has to do with the strength of the
interactions that keep the inert singlinos in equilibrium which depend heavily
on the mass of the $Z'$-boson mass.
If the inert singlinos do decouple in this range, this leads to a definite
prediction for $N_\eff$.
We shall explain why the inert singlinos
typically
decouple in this temperature range in the next subsection. For now we derive the
value of $N_\eff$ in this scenario as an example.

We shall use the subscript $0$ to denote quantities at some temperature $T_0$
below the electron mass and the subscript $e$ to denote quantities at some
temperature $T_e$ above the electron mass and where all light neutrino
species are still in equilibrium. We shall use the subscript $s$ to denote
quantities at some still higher temperature $T_s$ above the colour transition
and the strange quark mass and where the inert singlinos are still in
equilibrium.

At $T_s$ the effective number of degrees of degrees of freedom contributing to
the expansion rate is
\be
g^\eff_s &=& g^\gamma + g^g + \nicefrac{7}{8}(g^e + g^\mu + g^u + g^d + g^s + 3g^\nu + 2g^{\tilde{\sigma}}) \nn\\
&=& 2+16+\nicefrac{7}{8}(4+4+12+12+12+6+4) = 65\nicefrac{1}{4}
\ee
and at $T_e$ it becomes
\be
g^\eff_e &=& 2 + \nicefrac{7}{8}
\left(6+4\left(\frac{T^{\tilde{\sigma}}_e}{T_e}\right)^4\right)
\ee
and at $T_0$ it becomes
\be
g^\eff_0 &=& 2 + \nicefrac{7}{8}
\left(6\left(\frac{T^\nu_0}{T_0}\right)^4+4\left(\frac{T^{\tilde{\sigma}}_0}{T_0}\right)^4\right),
\ee
taking into account that the neutrinos and inert singlinos now have different temperatures.

The entropy within a given volume $V$ due to a
relativistic $\rr{boson}, \rr{fermion}$ $i$ with
number of degrees of freedom $g^i$ is given by
\be
S^i &=& (1,\nicefrac{7}{8})g^i\frac{2\pi^2}{45}(T^i)^3V.
\ee
Since we are assuming that the inert singlinos decouple
before the strange quark threshold, in going from $T_s$ to $T_e$
we conserve the entropy in the co-moving volume separately for the
inert singlinos and for everything else. Specifically for the inert singlinos
\be
T_s^3V_s &=& (T^{\tilde{\sigma}}_e)^3V_e
\ee
and for everything else
\be
[g^\gamma + g^g + \nicefrac{7}{8}(g^e + g^\mu + g^u + g^d + g^s + 3g^\nu)]T_s^3V_s
&=& [g^\gamma + \nicefrac{7}{8}(g^e + 3g^\nu)]T_e^3V_e \nn\\
\Rightarrow\quad 61\nicefrac{3}{4}T_s^3V_s &=& 10\nicefrac{3}{4}T_e^3V_e.
\ee
This allows us to write
\be
\frac{T_s^3V_s}{T_e^3V_e} = \left(\frac{T_e^{\tilde{\sigma}}}{T_e}\right)^3
&=& \frac{10\nicefrac{3}{4}}{61\nicefrac{3}{4}} = \frac{43}{247}.
\ee
In going from $T_e$ to $T_0$ we conserve the entropy separately for the
neutrinos, for the inert singlinos again, and for everything else
\be
[g^\gamma + \nicefrac{7}{8}g^e]T_e^3V_e &=& g^\gamma T_0^3V_0, \\
T_e^3V_e &=& (T_0^\nu)^3V_0, \\
(T^{\tilde{\sigma}}_e)^3V_e &=& (T_0^{\tilde{\sigma}})^3V_0.
\ee
This gives us
\be
\left(\frac{T_0^{\nu}}{T_0}\right)^3 &=& \frac{g^\gamma}{g^\gamma + \nicefrac{7}{8}g^e} = \frac{4}{11}
\ee
and
\be
\left(\frac{T_0^{\tilde{\sigma}}}{T_0}\right)^3 &=& \frac{43}{247}\frac{g^\gamma}{g^\gamma + \nicefrac{7}{8}g^e}
= \frac{43}{247}\frac{4}{11}.
\ee
In this case the effective number of neutrinos contributing to the expansion
rate prior to nucleosynthesis (at $T_0$) is then
\be
N_\eff &=& 3 + 2\left(\frac{43}{247}\right)^{4/3} \approx 3.194.
\ee

\subsection{The Inert Singlino Decoupling Temperature}
The light neutrinos are kept in equilibrium via their electroweak interactions.
For all the light neutrinos there are the following tree-level diagrams:
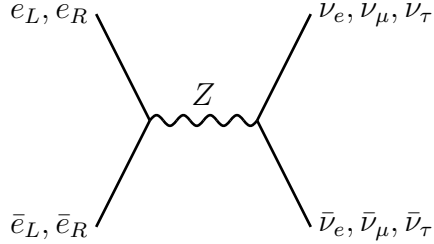
\begin{center}
\begin{picture}(120,120)(-60,-60)
\SetWidth{1}
\Line(-40,40)(-20,0)
\Line(-20,0)(-40,-40)
\Photon(-20,0)(20,0){2}{4}
\Line(40,40)(20,0)
\Line(20,0)(40,-40)
\Text(-43,40)[r]{$e_L,e_R$}
\Text(-43,-40)[r]{$\bar{e}_L,\bar{e}_R$}
\Text(0,10)[c]{$Z$}
\Text(43,40)[l]{$\nu_e,\nu_\mu,\nu_\tau$}
\Text(43,-40)[l]{$\bar{\nu}_e,\bar{\nu}_\mu,\bar{\nu}_\tau$}
\end{picture}
\end{center}
For the electron neutrinos there is also the following additional diagram:
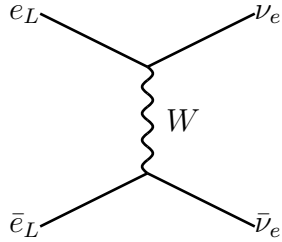
\begin{center}
\begin{picture}(120,120)(-60,-60)
\SetWidth{1}
\Line(-40,40)(0,20)
\Line(0,20)(40,40)
\Photon(0,20)(0,-20){2}{4}
\Line(-40,-40)(0,-20)
\Line(0,-20)(40,-40)
\Text(-40,40)[r]{$e_L$}
\Text(-40,-40)[r]{$\bar{e}_L$}
\Text(40,40)[l]{$\nu_e$}
\Text(40,-40)[l]{$\bar{\nu}_e$}
\Text(7,0)[l]{$W$}
\end{picture}
\end{center}
We express the cross-section for processes relevant for keeping
muon and $\tau$ neutrinos in equilibrium as
\be
\langle\sigma_{\nu_\mu,\nu_\tau}v\rangle &=& k_2\frac{T^2}{m_Z^4}\frac{(\nicefrac{5}{3})^2g_1^4}{s_W^4}X^4,
\ee
where $k_2$ is another constant and
\be
X^4 = ((-\nicefrac{1}{2}+s_W^2)(\nicefrac{1}{2}))^2 + \nicefrac{1}{4}s_W^4 \approx 0.031.
\ee
Note that
using the GUT normalised $U(1)_Y$ gauge coupling $g_1$ we have
\be
\frac{g_2}{c_W} &=& \sqrt{\frac{5}{3}}\frac{g_1}{s_W}.
\ee
The cross-section for electron neutrinos with their extra diagram is then
\be
\langle\sigma_{\nu_e}v\rangle &=& k_2\frac{T^2}{m_Z^4}\frac{(\nicefrac{5}{3})^2g_1^4}{s_W^4}Y^4,
\ee
where
\be
Y^4 = ((\nicefrac{1}{2}+s_W^2)(\nicefrac{1}{2}))^2 + \nicefrac{1}{4}s_W^4 \approx 0.147.
\ee
We express the number densities of all Weyl fermions still in equilibrium with
the photon as
\be
n^{e_L} = n^{e_R} = n^{\mu_L} = n^{\mu_R} = n^{\nu_e} = n^{\nu_\mu} = n^{\nu_\tau} &=& k_3T^3
\ee
and the expansion rate is given by
\be
H &=& k_1\sqrt{g_e^\eff}T^2.
\ee
The neutrino decoupling temperature $T^{\nu}$ can then be approximated by
\be
\langle\sigma_\nu v\rangle n^\nu &=& H \\
\Rightarrow\quad (T^{\nu_\mu,\nu_\tau})^3 &=& K\sqrt{g_e^\eff}m_Z^4\frac{s_W^4}{(\nicefrac{5}{3})^2g_1^4}\frac{1}{X^4}, \\
(T^{\nu_e})^3 &=& K\sqrt{g_e^\eff}m_Z^4\frac{s_W^4}{(\nicefrac{5}{3})^2g_1^4}\frac{1}{Y^4},
\ee
with $K = k_1/k_2k_3$. A more detailed calculation finds that in the SM
(with only neutrinos, electrons and photons contributing to $g_e^\eff$)
$T^{\nu_\mu,\nu_\tau} \approx 3.7$ MeV and
$T^{\nu_e} \approx 2.4$ MeV, the muon and $\tau$ neutrinos decoupling earlier.

At temperatures above the strange quark mass the processes relevant for keeping
the inert singlinos in equilibrium are as follows:
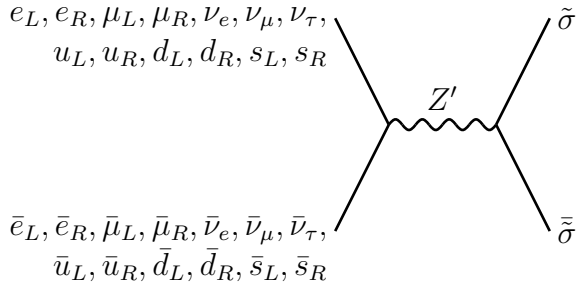
\begin{center}
\begin{picture}(120,120)(-60,-60)
\SetWidth{1}
\Line(-40,40)(-20,0)
\Line(-20,0)(-40,-40)
\Photon(-20,0)(20,0){2}{4}
\Line(40,40)(20,0)
\Line(20,0)(40,-40)
\Text(-43,40)[r]{$e_L,e_R,\mu_L,\mu_R,\nu_e,\nu_\mu,\nu_\tau,$}
\Text(-43,27)[r]{$u_L,u_R,d_L,d_R,s_L,s_R$}
\Text(-43,-40)[r]{$\bar{e}_L,\bar{e}_R,\bar{\mu}_L,\bar{\mu}_R,\bar{\nu}_e,\bar{\nu}_\mu,\bar{\nu}_\tau,$}
\Text(-43,-53)[r]{$\bar{u}_L,\bar{u}_R,\bar{d}_L,\bar{d}_R,\bar{s}_L,\bar{s}_R$}
\Text(0,10)[c]{$Z'$}
\Text(43,40)[l]{$\tilde{\sigma}$}
\Text(43,-40)[l]{$\bar{\tilde{\sigma}}$}
\end{picture}
\end{center}
The part of the $Z'$ current illustrating the relevant $U(1)_N$ charges is
\be
J^\mu_{Z'} &=& \left(\begin{array}{cccccc}
\bar{L} & \bar{e}_R & \bar{Q} & \bar{u}_R & \bar{d}_R & \bar{\tilde{\sigma}}
\end{array}\right)\gamma^\mu\frac{1}{\sqrt{40}}\left(\begin{array}{c}
(2)L \\ (1)e_R \\ (1)Q \\ (1)u_R \\ (2)d_R \\ (5)\tilde{\sigma}
\end{array}\right)
\ee
and the total cross-section taking into account all of these diagrams is then
(neglecting the small $Z$-$Z'$ mixing)
\be
\langle\sigma_{\tilde{\sigma}}v\rangle &=& k_2\frac{T^2}{m_{Z'}^4}2g_1^4\frac{Z^4}{(40)^2},
\ee
where
\be
Z^4 &=& (5)^2[2(2)^2 + 2(1)^2 + 3(1)^2 + 3(1)^2 + 6(1)^2 + 6(2)^2 + 3(2)^2] = 1450,
\ee
leading to an approximate singlino decoupling temperature of
\be
(T^{\tilde{\sigma}})^3 &=& K\sqrt{g_s^\eff}m_{Z'}^4\frac{1}{g_1^4}\frac{(40)^2}{Z^4} \\
\Rightarrow\quad \left(\frac{T^{\tilde{\sigma}}}{T^{\nu_e}}\right)^3
&=& \sqrt{\frac{g_s^\eff}{g_e^\eff}}\left(\frac{m_{Z'}}{m_Z}\right)^4\frac{(40)^2(\nicefrac{5}{3})^2}{s_W^4}\frac{Y^4}{Z^4}.
\ee
The only unknown variable here affecting the inert singlino decoupling
temperature is then the $Z'$ mass $m_{Z'}$. Rearranging we find
\be
m_{Z'} &\approx& m_Z \left(\frac{T^{\tilde{\sigma}}}{6.60 \mbox{~MeV}}\right)^{3/4}.
\ee

\subsection{$N_\eff$ in the \ezssm}
We now check which values of $m_{Z'}$ are consistent with our assumption
that the inert singlinos decouple at a temperature between the strange and
charm quark masses. For $T^{\tilde{\sigma}} < m_c$ we find that we require
$m_{Z'} < 4700$~GeV. For
$m_{Z'} \sim 1000$~GeV the situation is slightly more complicated. Firstly
the temperature of the QCD phase transition is not accurately known and
secondly the effective number of degrees of freedom is decreased by so much
after the QCD phase transition that even if the inert singlinos were decoupled
beforehand the universe may be expanding slowly enough afterwards that
they could come back into equilibrium. After checking a range of scenarios
we find that for $1300\mbox{~GeV}\lesssim m_{Z'} < 4700$~GeV our
value of $N_\eff = 3.194$ is valid. For $m_{Z'} \lesssim 950$~GeV the inert
singlinos decouple at a temperature above the muon mass, but below the pion
mass leading to a larger prediction of $N_\eff = 4.373$.
The current experimental limit in the \ezssm
is $m_{Z'} > 892$~GeV~\cite{Accomando:2010fz},
so at the time of writing it is possible that the $Z'$-boson is light enough 
to predict $N_\eff = 4.373$.
For $Z'$ masses in between these ranges the value of $N_\eff$
depends on the details of the QCD phase transition, but is somewhere between
these predictions. For inert singlinos decoupling above the pion mass,
but after the
QCD phase transition we have $N_\eff = 4.065$. All of these values are within
the 2-sigma measured range $\neff = 3.80^{+0.80}_{-0.70}$ and closer to the
central value than the SM result $\neff = 3$.

\section{Benchmark Points}
\label{benchmarks}
In the following tables we present three benchmark points in the c\ezssmSTOP
For all three points we fix $\lambda_{322}=0.1$ and $\lambda_{321} =
\lambda_{312} = 0.0001$ at the EWSB scale. For the $\dd{Z}_2^H$-breaking
couplings we also fix $\lambda_{332} = \lambda_{323} = 0.012$ and
$\lambda_{331} = \lambda_{313} = 0.005$ at the EWSB scale. At the GUT scale
we fix $\kappa_{333} = \kappa_{322} = \kappa_{311}$ and $\kappa_{3ij} = 0$ for
$i \ne j$.
The lightest (SM-like) Higgs mass is calculated to second loop order.

\begin{table}
\begin{center}\begin{tabular}{r|ccc|}
\hline
Benchmark & 1 & 2 & 3 \\\hline
$\tan(\beta)$ & 30 & 10 & 3 \\
$s$ [TeV] & 5 & 4.4 & 5.5 \\
$\lambda_{333}$ @ GUT scale & -0.3 & -0.37 & -0.4 \\
$\lambda_{322}$ @ EWSB scale & 0.1 & 0.1 & 0.1 \\
$\lambda_{311}$ @ EWSB scale & 0.0293 & 0.0403 & 0.0399 \\
$\kappa_{3ii}$ @ GUT scale & 0.18 & 0.18 & 0.23 \\\hline
$M_{1/2}$ [GeV] & 590 & 725 & 908 \\
$m_0$ [GeV] & 1533 & 454 & 1037 \\
$A$ [GeV] & 1375 & 1002 & 413 \\\hline
\end{tabular}\end{center}
\caption{The GUT scale parameters of the three benchmark points.}\end{table}

\begin{table}
\begin{center}\begin{tabular}{r|ccc|}
\hline
Benchmark & 1 & 2 & 3 \\\hline
$\mu$ [GeV] & -1086.7 & -1189.5 & -1405.5 \\
$\lambda_{322}s/\sqrt{2}$ [GeV] & 353.55 & 331.13 & 388.91 \\
$\lambda_{311}s/\sqrt{2}$ [GeV] & 103.59 & 125.38 & 155.17 \\\hline
$\tilde{N}_1$ mass [GeV] & 94.07 & 114.49 & 143.50 \\
$\tilde{N}_2$ mass [GeV] & -105.12 & -126.45 & -156.57 \\
$\tilde{N}_3$ mass [GeV] & 105.14 & 126.47 & 156.62 \\
$\tilde{N}_4$ mass [GeV] & 167.05 & 203.19 & 255.47 \\
$\tilde{N}_5$ mass [GeV] & -353.77 & -311.29 & -389.12 \\
$\tilde{N}_6$ mass [GeV] & 353.78 & 311.30 & 389.13 \\
$\tilde{N}_7$ mass [GeV] & -1092.5 & -1194.5 & 1409.6 \\
$\tilde{N}_8$ mass [GeV] & 1093.3 & 1194.8 & -1411.2 \\
$\tilde{N}_9$ mass [GeV] & -1803.2 & -1572.3 & -1964.7 \\
$\tilde{N}_{10}$ mass [GeV] & 1899.7 & 1688.7 & 2109.9 \\\hline
$\tilde{C}_1$ mass [GeV] & 105.04 & 126.41 & 156.52 \\
$\tilde{C}_2$ mass [GeV] & 167.05 & 203.19 & 255.46 \\
$\tilde{C}_3$ mass [GeV] & 353.78 & 311.30 & 389.13 \\
$\tilde{C}_4$ mass [GeV] & -1094.4 & -1196.1 & -1411.3 \\\hline
$m_{Z'}$ [GeV] & 1850.4 & 1628.4 & 2035.4 \\
$N_\eff$ & 3.194 & 3.194 & 3.194 \\
$\Omega_{\rr{CDM}}h^2$ & 0.112 & 0.107 & 0.102 \\
$\Upsilon$ & $1.1 \times 10^8$ & $2.3 \times 10^8$ & $2.3 \times 10^8$ \\
$\sigma_{\rr{SI}}$ [cm$^2$] & $4.9 \times 10^{-48}$ & $2.5\times 10^{-48}$ & $1.2 \times 10^{-48}$ \\\hline
\end{tabular}\end{center}
\caption{The low energy neutralino and chargino masses, and associated parameters.
The dark matter candidate is the lightest neutralino $\tilde{N}_1$, which is predominantly bino.
There is a nearby
pair of inert neutral Higgsinos $\tilde{N}_2$, $\tilde{N}_3$ and a chargino $\tilde{C}_1$
into which $\tilde{N}_1$ inelastically scatters
during freeze-out, resulting in the correct relic density $\Omega_{\rr{CDM}}h^2$ shown.
The predicted values of $m_{Z'}$ and $N_\eff$ are also shown,
as is the spin-independent $\tilde{N}_1$ direct detection cross-section $\sigma_{\rr{SI}}$.
\label{dmtable}}\end{table}

\begin{table}
\begin{center}\begin{tabular}{r|ccc|}
\hline
Benchmark & 1 & 2 & 3 \\\hline
$h_1$ mass [GeV] & 122.2 & 114.6 & 115.3 \\
$h_2$ mass [GeV] & 1145 & 987.1 & 1522 \\
$h_3$ mass [GeV] & 1890 & 1664 & 2080 \\
$H^\pm$ mass [GeV] & 2106 & 1396 & 1675 \\
$A^0$ mass [GeV] & 2103 & 1393 & 1673 \\\hline
$m_{S_2},m_{S_1}$ [GeV] & 1547 & 518 & 1084 \\
$m_{H_{d2}},m_{H_{d1}}$ [GeV] & 1567 & 611 & 1156 \\
$m_{H_{u2}},m_{H_{u1}}$ [GeV] & 1561 & 599 & 1146 \\\hline
$m_{\tilde{D}_3}$ [GeV] & 1483 & 503 & 1794 \\
$m_{\tilde{D}_2},m_{\tilde{D}_1}$ [GeV] & 1443 & 493 & 1775 \\
$m_{\tilde{\bar{D}}_3}$ [GeV] & 2864 & 2321 & 3065 \\
$m_{\tilde{\bar{D}}_2},m_{\tilde{\bar{D}}_1}$ [GeV] & 2840 & 2318 & 3052 \\\hline
$m_{\tilde{t}_1}$ [GeV] & 1122 & 625.3 & 1110 \\
$m_{\tilde{c}_1}, m_{\tilde{u}_1}$ [GeV] & 1817 & 1774 & 1707 \\
$m_{\tilde{t}_2}$ [GeV] & 1470 & 1069 & 1546 \\
$m_{\tilde{c}_2}, m_{\tilde{u}_2}$ [GeV] & 1838 & 1224 & 1761 \\
$m_{\tilde{b}_1}$ [GeV] & 1434 & 1009 & 1512 \\
$m_{\tilde{s}_1}, m_{\tilde{d}_1}$ [GeV] & 1840 & 1226 & 1763 \\
$m_{\tilde{b}_2}$ [GeV] & 1748 & 1265 & 1818 \\
$m_{\tilde{s}_2}, m_{\tilde{d}_2}$ [GeV] & 1907 & 1278 & 1820 \\
$m_{\tilde{\tau}_1}$ [GeV] & 1500 & 718.8 & 1259 \\
$m_{\tilde{\mu}_1}, m_{\tilde{e}_1}$ [GeV] & 1655 & 731.3 & 1261 \\
$m_{\tilde{\tau}_2}$ [GeV] & 1708 & 949.2 & 1473 \\
$m_{\tilde{\mu}_2}, m_{\tilde{e}_2}$ [GeV] & 1775 & 952.8 & 1474 \\
$m_{\tilde{\nu}_\tau}$ [GeV] & 1705 & 945.6 & 1472 \\
$m_{\tilde{\nu}_\mu}, m_{\tilde{\nu}_e}$ [GeV] & 1774 & 949.5 & 1472 \\\hline
$m_{\tilde{g}}$ [GeV] & 541.3 & 626.9 & 787.7 \\\hline
\end{tabular}\end{center}
\caption{The remaining particle spectrum in the standard notation.}\end{table}

We have chosen three points with quite different values of $\tan(\beta)$---30, 10 and 3.
This illustrates the fact that $\tan(\beta)$ can be quite low in this model since
the SM-like Higgs mass is not constrained to be less than $m_Z|\cos(2\beta)|$
at tree-level as it is in the MSSM.

The mass of the bino DMC $\tilde{N}_1$ is not directly constrained to be above above 100~GeV.
However, the lightest pseudo-Dirac inert Higgsino neutralinos $\tilde{N}_2$ and $\tilde{N}_3$ are almost
degenerate with the lightest inert Higgsino chargino $\tilde{C}_1$ and therefore these are constrained to
heavier than 100~GeV in order to be consistent with LEP constraints. Furthermore the
thermal relic DM
scenario outlined in Section~\ref{dm} requires $\tilde{N}_2$ and $\tilde{N}_3$ not to be
too much more massive than $\tilde{N}_1$. In practice the $\tilde{N}_1$ is predominantly bino and its
mass cannot be much less
than 100~GeV. In Benchmark~1, for example, it is 94~GeV.

Requiring such values for the low energy bino mass $M_1$ and requiring consistent
electroweak symmetry breaking in practice means that the SM-singlet VEV $s$ cannot be
too low. This in turn means that the $Z'$ mass is always quite a bit above the experimental
limits, more than about 1.5~TeV. In these benchmarks from the constrained scenario the
effective number of neutrinos contributing to the expansion rate of the universe prior to
BBN $N_\eff$ therefore takes on the lower value calculated in Section~\ref{neff}---around 3.2.
This is more consistent with data than the SM prediction.

In all benchmark points $\tilde{N}_4$ and $\tilde{C}_2$ are predominantly wino. $\tilde{N}_5$,
$\tilde{N}_6$ and $\tilde{C}_3$ are predominantly made up of
the rest of the inert Higgsinos states, with masses around $\lambda_{322}s/\sqrt{2}$, whereas
$\tilde{N}_7$, $\tilde{N}_8$ and $\tilde{C}_4$ are predominantly made up of
the active Higgsinos states, with masses around $\mu$. $\tilde{N}_9$ and $\tilde{N}_{10}$
are mostly superpositions of the active singlino and bino$'$.

The fact that $\Upsilon \gg 1$ indicates that the inert Higgsino components in the
predominantly bino state $\tilde{N}_1$, though small, are large enough such that
processes involving $\tilde{N}_1$ up-scattering off of a SM particle into $\tilde{N}_2$
happen overwhelmingly more often than neutralino (co-)annihilation processes. In this
way the ratios of the number densities of these particles are able to maintain their
equilibrium values.

The spin-independent
DMC-nucleon cross-section $\sigma_{\rr{SI}}$, as estimated using the results
of Choi \textit{et al.}~\cite{Choi:2000kh}, is quite small for these benchmarks,
and is not currently detectable by direct detection experiments.
This is due to the predominantly bino nature of the DMC,
and the large squark masses.

\section{Conclusions}
\label{conclusions}
The question of dark matter in the \essm illustrates the interesting diversity
of possibilities that that can arise once one goes beyond the MSSM.
The difficulty in making the inert singlinos predicted by this model
much heavier than 50~GeV makes them
natural dark matter candidates, but also led to a very tightly constrained scenario
in which the inert LSP (essentially a mixture of the inert singlinos and inert Higgsinos)
is now severely challenged by the most recent XENON100 analysis of 100.9
days of data. Moreover
the tightly constrained parameter space makes it practically impossible 
for such a scenario to be consistent with having universal soft mass parameters.

In this paper we have discussed a new variant of the $E_6$SSM, 
called the EZSSM,
which involves a novel scenario for Dark Matter in which 
the dark matter candidate is predominantly bino with a mass 
close to or above 100~GeV which is fully consistent with XENON100.
A successful relic density is achieved via
its inelastic scattering into nearby heavier inert Higgsinos during the time of thermal freeze-out.
The model
also predicts two massless inert singlinos which contribute to the
number of effective neutrino species at the time of Big Bang Nucleosynthesis, depending
on the mass of the $Z'$-boson which keeps them in equilibrium. For example for 
$m_{Z'} > 1300$~GeV we find $N_\eff \approx 3.2$.

We have studied a few benchmark points
in the constrained \ezssm with massless inert singlinos to illustrate this new scenario.
The benchmark points show that it is easy to find consistent points which satisfy the correct
relic abundance as well as all other phenomenological constraints. The points also show that 
the typical $Z'$ mass is expected to be around 2~TeV, with the gluino having a mass
around 500--800~GeV and squarks and sleptons typically having masses around 1--2~TeV.
The direct detection spin-independent cross-sections $\sigma_{\rr{SI}}\sim {\rm few}\times 10^{-48}$~cm$^2$
are well below current sensitivities.
\section*{Acknowledgements}
We would like to thank Peter Athron for donating his code on the c\essmSTOP
JPH is thankful to the STFC for providing studentship funding.
SFK acknowledges partial support from the STFC Rolling Grant ST/G000557/1.
\bibliographystyle{JHEP}
\bibliography{library}

\end{document}